\documentclass[usenatbib,twocolumn,useAMS]{mn2e}
\usepackage{epsfig}

\def\msun{{\rm M_{\odot}}}

\title [Photoevaporation of Discs ]
{The Photoevaporation of  Discs Around Young Stars in
Massive Clusters} 
\author[C.J.~Clarke]{C.J.~Clarke$^1$\\
$^1$Institute of Astronomy, Madingley Rd, Cambridge, CB3 0HA, UK}
\date{Submitted: July 2103}

\pagerange{\pageref{firstpage}--\pageref{lastpage}} \pubyear{2003}

\begin{document}
\def\lta{\mathrel{\spose{\lower 3pt\hbox{$\mathchar"218$}}
     \raise 2.0pt\hbox{$\mathchar"13C$}}}
\def\gta{\mathrel{\spose{\lower 3pt\hbox{$\mathchar"218$}}
     \raise 2.0pt\hbox{$\mathchar"13E$}}}
\def\Msun{{\rm M}_\odot}
\def\msun{{\rm M}_\odot}
\def\Rsun{{\rm R}_\odot}
\def\Lsun{{\rm L}_\odot}
\def\19{GRS~1915+105}
\label{firstpage}
\maketitle

\begin{abstract}

We present models in which the photoevaporation of discs around
young stars by an external 
ultraviolet source (as computed by Adams et al 2004) is coupled with the 
internal viscous evolution of the discs. These models are applied to the
case of the Orion Nebula Cluster,  where the presence of a strong
ultraviolet field from the central OB stars, together with
a detailed  census of circumstellar
discs and photoevaporative flows, is well established. 
In particular we investigate the
constraints that are placed on the initial disc properties in the ONC
by the twin requirement that most stars possess a disc on a scale of
a few A.U., but that only a minority ($< 20 \%$) are resolved by HST at
a scale of $50$ A.U.. We find that these requirements place very weak
constraints on the initial radius distribution of circumstellar discs:
the resulting size distribution readily forgets the initial radius
distribution, owing to the strong positive dependence of the photoevaporation
rate on disc radius. Instead, the scarcity of large discs reflects the
relative scarcity of initially massive discs (with mass $> 0.1 M_\odot$).
The ubiquity of discs on a small scale, on the other hand, mainly
constrains the timespan over which the discs have been exposed to
the ultraviolet field ($< 2 $Myr). We argue that the discs that are resolved
by HST represent a population of discs in which self-gravity was important
at the time that the dominant central OB star switched on, but that, according to
our models, self-gravity is unlikely to be important in these discs at the
present time. We discuss the implications of our results for the
so-called proplyd lifetime problem.    
\end{abstract}


\section{Introduction}

  The Orion Nebula Cluster (ONC) provides a unique opportunity to measure the 
size distribution of discs around young stars. Whereas in other 
regions, resolved discs are observed in emission - and therefore the deduced sizes are sensitive to factors such as the disc temperature
profile or the depletion or dissociation of molecular tracers - 
the bright nebular background of the ONC allows discs to be
observed {\it in silhouette}. In this case, the edges of
observed discs represent uniform contours in optical extinction
and hence, assuming standard dust:gas ratios, uniform contours
of gas column density. Silhouette discs therefore provide the best
opportunity to measure disc size distributions with the fewest
input assumptions about conditions in the disc. 

  Disc sizes are of interest inasmuch as they provide information
about the initial angular momentum of the natal gas (Burkert
and Bodenheimer 2000), about the
possibility of disc truncations by dynamical interactions 
(Armitage and Clarke 1997, Bate, Bonnell and Bromm 2003)
and about the magnitude of the viscosity driving disc
spreading (Hartmann et al 1998). Disc sizes evidently also
limit the range of separations over which planets can
potentially form. Most importantly, the lifetime of an isolated
disc is determined by the viscous timescale at its outer edge, 
which is almost certainly an
increasing function of disc outer radius. The disc size
distribution is thus intimately linked with the large range of
lifetimes of circumstellar discs and probably holds the
key to understanding the mixture of disc possessing and
discless systems seen in stars of a given age
(Armitage, Clarke and Palla 2003, Alexander and Armitage 2006,
Dullemond et al 2006). 

  However, the ONC is {\it not} a good environment in which to use
disc sizes as a diagnostic of viscous spreading, since disc sizes
in Orion are shaped, in addition, by photoevaporation by the FUV
continuum of the cluster's central OB stars
(particularly its most massive member, the O6 star
$\theta_1 C$). Direct evidence for the effect of photoevaporation
on disc sizes is provided by the fact that relatively
few silhouette discs in Orion have been resolved
by HST at a size scale of $50$ A.U. (McCaughrean and Rodmann, in
preparation), whereas in Taurus-Auriga
(which lacks massive stars), the majority of discs are resolved
by submillimetre studies at a scale of $100$ A.U. or more
(Kitamura et al 2002, Andrews and Williams 2007). In Orion,  photoevaporative
flows  are manifest as {\it proplyds}, cometary shaped ionisation fronts that are
associated with a number of stars in the inner regions of the ONC, both those
with and without resolved silhouette discs
(O'Dell et al 1993, O'Dell and Wen 1994, O'Dell and Wong 1996, McCaughrean and
O'Dell 1996, Bally et al 1998a), Bally et al 2000).
In the standard, and highly successful, models for proplyds, the
FUV radiation field in Orion creates a layer of warm outflowing gas,
fed by the disc, into which ionising photons can only
penetrate to within a couple of disc radii of the central star
(e.g.  Johnstone et al 1998, St\"orzer and Hollenbach 1999).
The mass loss rate deduced from these models - independently
confirmed by spectroscopic measurements (Henney \& O'Dell 1999)
- are high ($\sim 10^{-7} M_\odot$ yr$^{-1}$) thus raising
the issue of the longevity of circumstellar discs in this
environment.

  In a recent elaboration of disc photoevaporation theory,  Adams
et al (2004) have constructed models in which discs lose mass mainly
through their outer rims. These models  couple hydrodynamic flow
equations to the complex relationship between temperature and column density
provided by detailed PDR modeling. In essence, the flow resembles
a Parker wind solution in that material leaves the disc edge in a
subsonic flow and is accelerated by thermal pressure gradients to
sonic velocities at a location close to the point where the flow
becomes marginally unbound.  Unlike the classic Parker wind solution,
however,  this flow is not isothermal - material is injected
into the base of the PDR at a few hundred K and is heated to a
few thousand K as it flows away from the disc edge, due to
its increasing exposure to the incident ultraviolet radiation
field. The results of Adams et al suggest that although discs
experience large mass loss rates 
if they are extended to $> 100$ A.U., this rate drops strongly
for more compact discs. [For example,
they find mass loss rates of $\sim 10^{-8} M_\odot$ yr$^{-1}$ for
discs of radius $40$ A.U. and $\sim 10^{-9} M_\odot$ yr$^{-1}$ for
outer radius of $20$ A.U., under conditions where the mass loss
rate exceeds ($ 10^{-7} M_\odot$ yr$^{-1}$) for large
($> 100$ A.U.) discs]. The reason for this steep dependence
on disc outer radius is that material at the disc outer edge is more
tightly bound in the case of more compact discs. It therefore
has to be heated to higher temperatures to initiate the flow, and
these higher temperatures can only be achieved if the flow is
lower in density. 

  In this paper we use the apparatus set up by
Adams et al to study the evolution of discs that are
subject to both photoevaporation and internal viscous
evolution.
Our time dependent  models employ
a variety of assumptions about the initial mass and radius
of the discs and we examine how these parameters are constrained by
two key observations: the observed high fraction of stars with
discs on the scale of an A.U. or less , and the much lower fraction possessing
discs resolved on a scale of $50$ A.U. or more. Section 2 sets out the
method and assumptions, Section 3 describes the nature of the evolution
and Section 4 examines what constraints are placed on the initial
masses and radii of discs in the ONC. Section 5 discusses the results
in relation to the proplyd lifetime problem and the role of
self-gravity in circumstellar discs. Section 6 summarises the
conclusions.

\section{Method}
\subsection{Treatment of gas}

  We model the disc as a viscous accretion disc subject to
photoevaporative mass loss at its outer edge. 
For the viscous evolution of the disc we follow a number of authors
(e.g. Hartmann et al 1998, Clarke et al 2001, Armitage et al
2003) in assuming that the disc viscosity is proportional
to radius in the disc. Such a parameterisation is equivalent
to a disc with a constant $\alpha-$ viscosity (Shakura \& Sunyaev
1973) and a mid-plane temperature that declines with radius
as $\propto r^{-1/2}$, and implies that the steady state
surface density distribution declines as $r^{-1}$. (See
Hartmann et al 1998 for arguments in favour of such a viscosity
law).  

  A convenient aspect of viscosity laws that are power laws in radius
is that  there are similarity solutions for the disc evolution
(Lynden-Bell and Pringle 1974). For viscosity $\propto r$,
this takes the form:

\begin{equation}
\Sigma = {{M_d(0)}\over{2 \pi R_1^2 x}} exp\bigl(-{{x}\over{T}}\bigr) T^{-1.5}
\end{equation}

\noindent  where  $M_d(0)$ is the initial disc mass, $R_1$ is the initial
disc scaling radius,$x = r/R_1$ and $T=1+t/t_s$ where $t_s$ is the
viscous timescale ($r^2/3 \nu$) at $r=R_1$. Such a $\Sigma$ profile
implies that the accretion rate through the disc may be written: 

\begin{equation}
\dot M = {{1}\over{2}} {{M_d(0)}\over{t_s}} exp \bigl(-{{x}\over{T}}\bigr) T^{-1.5} \bigl(1 - {{2x}\over{T}}\bigr)
\end{equation}

This means that the accretion flow is inwards for $x < T/2$ and that the rate of outflow reaches a maximum at $x = 3T/2$, where the outward accretion
flow is equal to  its instantaneous value at small radii ($x << 1$) 
multiplied by 
a factor $2 e^{-1.5}$. The quantity $TR_1$ therefore controls the radius in the
disc separating inflowing and outflowing regions at any time.

  We evolve the viscous diffusion equation for such a disc using a standard
explicit finite difference method, equally spaced in $r^{1/2}$. At every
timestep, we locate the nominal disc edge , $r_{edge}$ (for the
purpose of locating the correct photoevaporative mass loss rate, $\dot M_{wind}
(r_{edge})$) by finding the grid cell where the column density is
closest to some low constant value ($\Sigma_{edge}$). (In practice,
since the disc develops a very steep density profile at its outer edge,
the location of $r_{edge}$ (and hence the value of $\dot M_{wind}(r_{edge})$)
is very insensitive to the value of $\Sigma_{edge}$, provided it is
sufficiently low). Having specified $\dot M_{wind}(r_{edge})$ (see below)
we progress inwards from the outer edge of the grid, removing a
fraction $1-\epsilon$ of the mass in each grid cell until a cell (i)
is
reached where the total mass removed this timestep exceeds $\dot M_{wind}(r_{edge}) \times dt$. Note that $\epsilon$ is not set equal to zero, in order to avoid
numerical difficulties with empty grid cells, but that again the evolution
is insensitive to $\epsilon$ provided $\epsilon$ is small.

  In order to specify $\dot M_{wind}(r_{edge})$, we apply prescriptions based
on the numerical and analytical results presented in Adams et al 2004.
Specifically, we assume (in units of $M_\odot$ yr$^{-1}$):

\begin{equation}
\dot M_{wind} = 5.1 \times 10^{-8} \biggl({r_{edge} \over 100 A.U.} \biggr) (a)
\end{equation}
\begin{equation}
\dot M_{wind} = 9.9 \times 10^{-8} \biggl({420 A.U. \over r_{edge} } \biggr)
exp \biggl({-210 A.U. \over r_{edge}}\biggr) (b)
\end{equation}
\begin{equation}
\dot M_{wind} = 1 \times 10^{-8} exp \biggl({(r_{edge}-40. A.U.) \over 27 A.U.}\biggr) (c) 
\end{equation}
\begin{equation}
\dot M_{wind} = 1 \times 10^{-8} exp \biggl({(r_{edge}-40. A.U.) \over 9 A.U.}\biggr) (d)
\end{equation}

(a) corresponds to  $r_{edge} > 100$ A.U., i.e.
the limit where the disc outer edge is comfortably
larger than the escape radius for photoevaporated gas, $r_{esc}$, (b)
to the  regime  ($60-100$ A.U.)
where Adams et al proposed an analytic prescription 
for the mass loss rate when the disc radius
is modestly less than $r_{esc}$ and (c) and (d)  (for $r_{edge}$
respectively in the range $40-60$ A.U. and $ < 40$ A.U.) are fits
to the numerical results of Adams et al.
represents a fit to
their numerical results in the case $r_{edge} << r_{esc}$. Figure 1 depicts our
mass loss prescriptions as a function of $r_d$ for 
$M_*= 1 M_\odot $ and ultraviolet radiation field intensity $G_0 = 3000$.
We note that the calculations of Adams et al
for the case of a higher radiation field ($G_0= 30000$), as is
appropriate to the very central regions of the ONC, imply wind mass loss
rates that exceed those above by only a modest factor ($\sim 2-3$). 
This relative insensitivity to the value of $G_0$ means that our
results (which are based on the assumption that a disc is subject to irradiation
by a constant flux over its lifetime) would not differ greatly if
one took into account the fact that a star's distance to the ionising
source is likely to change modestly over its orbital history
(see analysis in Scally and Clarke 2001).

\begin{figure}
\vspace{2pt}
\psfig{file=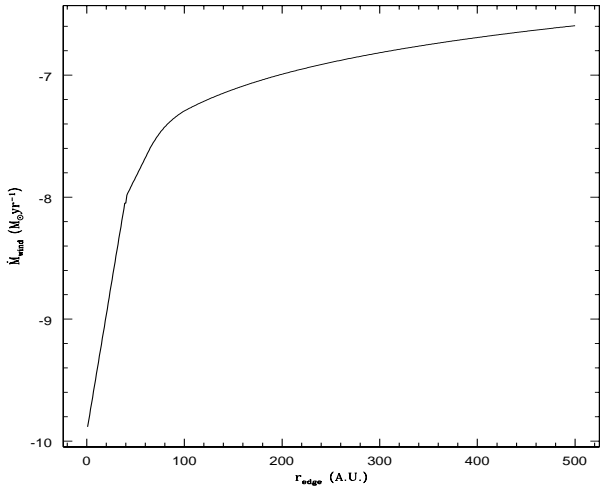,width=8.5cm,height=7.cm}
\caption{Assumed photoevaporative mass loss rate  as a function of outer
disc radius (equations (3)-(6)). Adapted from Adams et al 2004 for the
case $M_*= 1 M_\odot $ and ultraviolet radiation field intensity $G_0 = 3000$.}
\end{figure}
 
\subsection{Treatment of dust}

 In Section 4 we will compare the models with the observed distribution
of the  sizes of silhouette discs in the Orion Nebula Cluster. Since disc 
opacity is dominated by the dust, rather than gas, component, it is only
possible to relate the modeling of disc gas evolution (described above) to
disc sizes if we make some additional assumption about the evolution of
the dust.

 In what follows, we will simply assume that the dust follows the
evolution of the gas. Although this is unlikely to be true in detail
(see discussion below) we are motivated to do this because the disc
edges in our photoevaporating simulations are extremely sharp
and because the optical depth of the disc just inside this
edge is enormous ($ > 10^4$ for a standard grain mixture and
dust to gas ratio). This means that the gas disc's edge is well defined
and that the edge of the dust disc will also be well defined unless
the dust to gas ratio differs from standard values by a very large factor.

  A number of studies have however shown that it is likely that the
micron sized grains (which dominate the opacity at
optical wavelengths: Miyake and Nakagawa 1993) are well coupled to the
gas, both in the case of pure disc flows and in the case of
photoevaporation (Takeuchi et al 2005, Alexander and Armitage 2007). 
One possible route to creating a gas poor outer dust disc is if the
gas and associated small grains are photoevaporated, and then small
dust is replenished by collisions between residual large bodies in the
disc (Throop and Bally 2005). If the gas and dust did not trace
each other, however, one would not  expect to see the observed 
correlation between the size of discs embedded in ionised objects (proplyds)
and the offset of the ionisation fronts (Johnstone and Bally 1998, Vicente and
Alves 2005).

 In summary, although the study of dust evolution in discs subject
to external photoevaporation is an issue that could repay further
study, it is  unlikely that such studies would contradict 
our assumption here, i.e.
that the sizes of the gas discs derived from our models
can be directly compared with the observed sizes of silhouette discs.

 \section{Results: portrait of the evolution} 

  We find that, depending on the initial disc parameters, 
there are two qualitatively distinct evolutionary
patterns. If the initial accretion rate at the inner edge ($\dot M_{in}$)
is substantially larger than $\dot M_{wind}(R_1)$ (Figure 2), the disc will
self-adjust to a quasistatic configuration with radius ($> R_1$) such that the
outward viscous flow rate at $r_d$ just matches $\dot M_{wind}(r_d)$. 
This situation can be sustained over a time $T \sim (\dot M_{in}/\dot M_{wind}(r_d))^{2/3}$, over which the accretion rate on to the star declines by
viscous evolution to a level $\sim \dot M_{wind} (r_d)$. [Note that over
this stalled period, the density profile in the region of $r_d$ self-adjusts
so as to continue to deliver the required viscous flow of material
to re-supply the wind]. However, once the accretion rate on to the star
drops to below $\dot M_{wind} (r_d)$, the disc is unable to sustain
the viscous outflow at $r_d$ that is required to hold $r_d \sim$ constant.
Subsequently, the disc outer edge moves inwards on its instantaneous
viscous timescale. (Note that in ordinary viscous evolution, a fluid element at radius $r$ moves inwards on its viscous timescale ($t_\nu (r)$) but that
the surface density at $r$ evolves on the generally much longer timescale of the
viscous timescale at the outer edge, due to re-supply of material from
larger radius. In the present case, as fluid  moves in on a timescale $t_\nu$, the  
wind prevents re-supply of material from larger radius and hence $t_\nu$ is in
this case the timescale for the shrinkage of the disc's outer edge.) We
note that in this second evolutionary stage, the disc is  
drained by a mixture of accretion onto the star and  photoevaporation.
Since the viscous timescale scales as $r$ for our viscosity
prescription, one may understand that the resultant evolution during this stage
involves the disc moving in at roughly constant velocity ($\sim R_1/t_s$:
see Figure 4).
\begin{figure}
\vspace{2pt}
\psfig{file=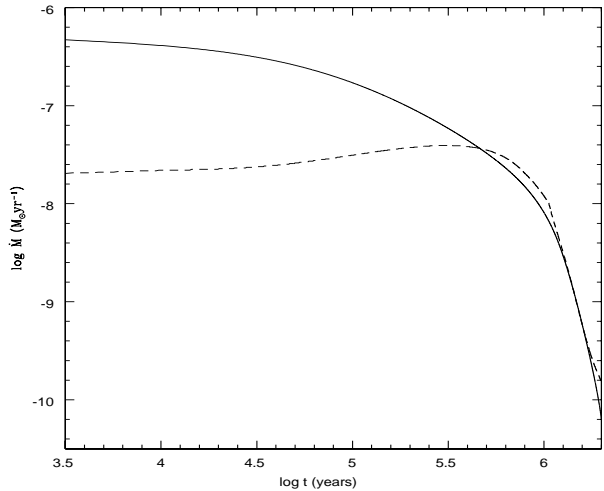,width=8.5cm,height=7.cm}
\caption{ Evolution of accretion rate onto the star (solid) and photoevaporation
rate at the disc outer edge (dashed) for a model with initial disc
mass $0.1 M_\odot$,  scaling radius $R_1=10$ A.U. and viscous timescale
$t_s = 1.1 \times 10^5$ years. }
\end{figure}

  The other type of evolution (illustrated in Figure 3, where the
disc mass is unchanged compared with Figure 1, but where the scaling
radius and viscous timescale is an order of magnitude larger
and the initial accretion rate hence an order of magnitude smaller) occurs in systems 
where $\dot M_{wind}(R_1) > \dot M_{in}$. In this case, the outer disc 
radius never stalls, but instead moves in on a timescale set by
the photoevaporative mass loss, i.e. by the fulfillment of the 
condition on $r_d(t)$ that:

\begin{equation} \int_0^t \dot M_{wind}(r_d'(t'))dt' = M_{di}(r>r_d(t)) \end{equation}

\noindent where $ M_{di}(r>r_d(t))$ is the initial disc mass outward of $r_d(t)$. (Note that
this condition assumes the photoevaporation of a static disc
and hence neglects the minor role of viscous evolution).During this
time, the accretion rate on to the central star remains roughly
constant, provided that the time to 
photoevaporate the disc to radius $r_d$ is less than the disc's
viscous timecale at $r_d$. Since, $\dot M_{wind}(r_d)$ is decreasing
monotonically as $r_d$ shrinks, whereas $\dot M_{in}$ is constant,
the two mass flow rates eventually become comparable and thereafter
the disc drains on a viscous timescale through a mixture of photoevaporation
and accretion. 

\begin{figure}
\vspace{2pt}
\psfig{file=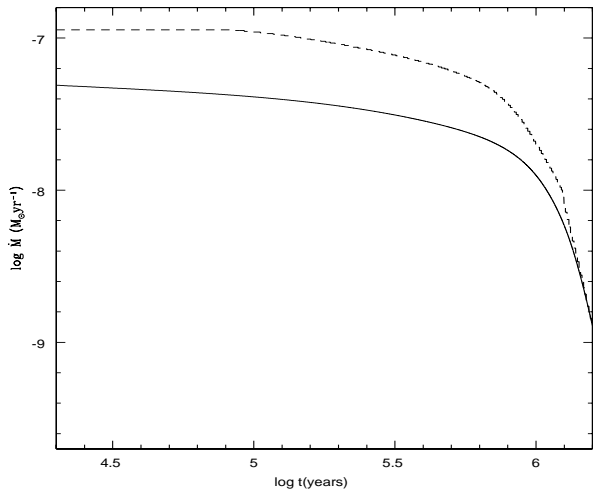,width=8.5cm,height=7.cm}
\caption{As Figure 2 for the case that the initial
disc mass is $0.1 M_\odot$, $R_1 = 100$ A.U. and $t_s = 1.1 \times
10^6$ years.}
\end{figure}

 The one to one correspondence between 
wind mass loss and disc outer edge radius (Figure 1) means that the dashed
curves in Figure 2 and 3 may be readily converted into plots of the
evolution of the disc outer edge. We display this evolution  (in linear
rather than logarithmic time units) in Figure 4 and draw attention to
the fact that the late time evolution of the two models (which share
the same value of initial disc mass but which differ in $R_1$ - and
hence initial accretion rate - by a factor of $10$) are very similar.
The models differ in their early evolution inasmuch as the extended
(low accretion rate) model (dashed in Figure 4) initially loses mass
mainly through photoevaporation, whereas the compact, high accretion rate
model initially drains by accretion. At late times, however, both
models converge on the same evolutionary path, with mass being
drained by accretion and photoevaporation in nearly equal measure. 
We note that the stalling of the outer edge in the compact case
(which can be inferred from the evolution of the wind mass loss rate
in Figure 2) is not very significant when plotted  against time
in linear units, since stalling occurs at early times when the disc
expands out to its maximum extent.  

\begin{figure}
\vspace{2pt}
\psfig{file=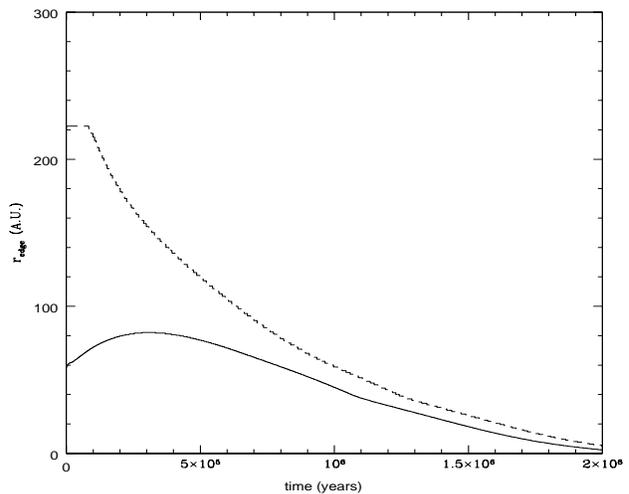, width=8.5cm,height=7.cm}
\caption{Disc radius evolution as a function of time for the models
depicted in Figures 2 and 3: $R_1 = 10$ A.U. (solid), $R_1=100$ A.U.
(dashed).} 
\end{figure}

\section{Comparison with observed disc size distribution}

 We will now attempt to constrain the permitted distribution of
initial disc parameters using observations of the observed size distributions
of discs in the ONC. The key observations are as follows:
(i) the majority of stars close to the centre of the ONC exhibit
ionised outflows: within the central  $0.15$ pc in projection, the  fraction of 
sources exhibiting such flows is $80 \%$ (Bally et al 2000), 
which is compatible with
{\it all} the stars within this region in three dimensions being subject to
such flows (Scally and Clarke 2001). 
(ii) out of a sample of $150$ sources exhibiting ionised
outflows, only $25$ contained discs that were resolvable by HST;
the minimum disc radius that could be extracted in this study varied over
the range $20-50$ A.U.: thus the fraction of discs with radius greater than
$50$ A.U. is considerably less than $20 \%$ 
( Rodmann 2002,
McCaughrean and Rodmann in preparation).\footnote{ Note that
this points to the importance of photoevaporation in pruning
discs in the ONC, since in Taurus, which lacks photoionising sources,
discs are considerably larger: according to Kitamura et al 2002, about half
the Classical T Tauri stars in Taurus were imaged in millimetre
emission at a scale of $100$ A.U. or more, whereas the recent
study of Andrews and Williams (2007) estimates a median disc radius in
Taurus of about $200$ A.U..} (iii) essentially {\it all}
sources exhibiting ionised outflows however exhibit thermal infrared
emission indicative of disc material at a radius of an A.U. or less
(Lada et al 2000).  From this we may conclude that 
stars in the core of the ONC are a) overwhelmingly likely to be subject
to photoevaopration (so that the models here will be applicable to them),
b) possess reservoirs of disc material  (as required by  models
of disc photoevaporation), but c) rather rarely (i.e. in less than $20 \%$ of
cases) exhibit discs on a scale of $50$ A.U. or more. For the subset
of ionised outflow  sources exhibiting resolved discs (i.e. $25$ objects)
the size distribution in the regime $r > 50$ A.U. may be described
as a power law of the form $n(r) \propto r^{-2}$ (McCaughrean and
Rodmann, private communication).   

  It is evidently impossible to use such information to reconstruct
the parameter distribution of initial disc radii and masses in the ONC,
owing to the considerable degeneracy in the problem. On the other
hand, we can derive interesting constraints by considering what are
the aspects of the observational data that are going to be hard to
reproduce by the sort of models we have constructed. We can get a hint
of this from Figure 4: evidently in both these models (which share
a common initial disc mass but in which the size and initial
accretion rate differ by an order of magnitude), the sources spend
$> 50 \%$ of their lives with radii greater than $50 $ A.U.. If star-disc
systems are created at a steady rate during the time that the 
photoionising source is `on', the observed fraction of systems  with
resolved discs then requires that sources should spend 
$< 20 \%$ of their lives with $r > 50 $ A.U., in clear contradiction
of the models.

  We have explored this further by considering a grid of models
parameterised by the initial disc scaling radius $R_1$ (see
Section 2) and initial disc mass ($M_d$). We assume a viscosity
law of the form $\nu \propto r$ and have normalised different
models such that the  viscous  
timescale at a given radius is the same for all models (in
terms of $\alpha $-viscosity theory, this implies that
$\alpha$ and the disc aspect ratio are the same in all models).
This means  that 
in different models, the initial
accretion rate always scales with the disc mass within a fixed radius.
Since the initial surface density distribution implies that the enclosed
mass within radius $r$ scales as $r/R_1$ (provided $r < R_1$), this means
that the initial accretion rate scales as $M_d/R_1$. We normalise
our models such 
that the viscous timescale at $10$ A.U.
is always $1.1 \times 10^5$ years. 

  For such a grid of models,  we then ask how long each disc spends with
$r > 50$ A.U. as a fraction of the total lifetime of the disc.
We find that there are essentially {\it no} credible models that
can match the data. In most models, the system spends considerably
longer with radius larger than $50$ A.U. than with  radius less
than $50$ A.U.. If we want to reverse the situation, then it is
difficult to change significantly 
the time that  the disc spends at small radii, given that
the erosion of the inner disc is always set roughly by its viscous
timescale (which we have not allowed to vary between models). 
On the other hand, we can change the balance of small and
large
discs (i.e. $<$ and $> 50$ A.U.) by ensuring that the 
outer disc is eroded extremely rapidly:
this is favoured
by low total disc mass and by large scaling radius.  
However, even in the case of very extreme parameters
(e.g.  initial disc mass of $3 \times 10^{-3} M_\odot$,
$R_1 = 200$ A.U.), the time spent by the system at small radius
never much exceeds the time spent at large radius; moreover, in this
case, all the `large discs' are at radii only slightly exceeding
$50$ A.U., whereas the data shows a smoothly declining
distribution out to an observed maximum of $> 150$ A.U..
\footnote {We have also experimented with changing the radial
dependence of the viscosity law,  considering also the
case that $\nu \propto r^{1.5}$ (i.e. where the steady state
surface density profile scales as $r^{-1.5}$). Such models
predict faster viscous evolution at large radius than the
standard models and also faster erosion by photoevaporation
at large radius, due to the lower disc mass at large radius.
Although these changes work in the direction of increasing
the fraction of compact  ($> 20$ A.U.) sources, they are still not
close to satisfying the observational constraints, and so we
consider that changing the visocity law is unlikely to
solve this problem.} 

  We then consider another scenario: i.e. that stars are not
being created at a steady rate but that instead star formation
ceased at the point that the photoionising source switched on,
at a time $t_{uv}$ ago. This assumption is not unreasonable given the
evidence from radio maps for the escape of ionised gas from the core
of the ONC (Wilson et al 1997), together with the theoretical
expectation that ionising radiation and stellar winds should clear
the star forming gas from the vicinity of young massive stars
(see, for example, Figure 16 of Dale et al 2005).   We find that if $t_{uv}$ were  much in excess
of $1.5$ Myr then the majority of discs would already have drained away due to
a mixture of photoevaporation and accretion, in contradiction to the
near ubiquity of sources with a near infrared excess in the core of the ONC.
\footnote{In the  case of a source switching on (and suppressing subsequent
star formation) , the distribution of disc
parameters should of course  be thought of as representing those of a population
of stars at the moment of ultraviolet switch on, rather than 
being necessarily a set of initial parameters}. 

\begin{figure}
\vspace{2pt}
\psfig{file=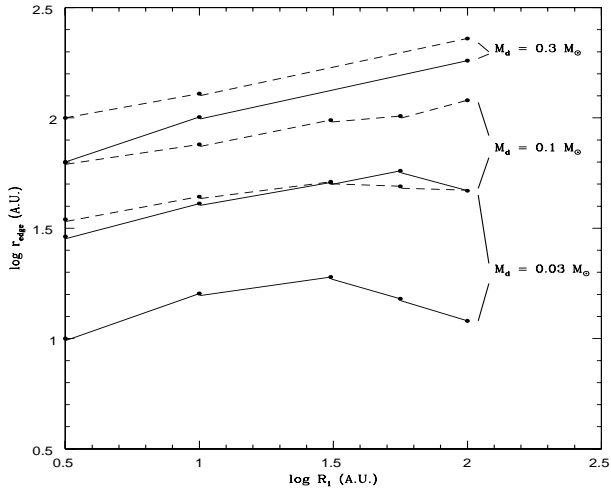,width=8.5cm,height=7.cm}
\caption{Disc radius as a function of initial disc scaling radius
($R_1$). The dashed (solid) lines link models of fixed initial disc
mass after $t_{uv}= 0.5$ ($1$) Myr of exposure to the photoionising source. Initial
disc masses for each pair of $t_{uv}$ values are listed on the right hand side of the plot.}
\end{figure}

 Figure 5 is a plot
of the disc edge radius as a function of $R_1$, with models of
the same initial disc mass being connected by  dashed and 
solid lines for the cases $t_{uv} = 0.5$ and $1$ Myr.
It is immediately evident that  although there is a mild
trend of increasing $r_{edge}$ with $R_1$ at a given $M_d$ and
$t_{uv}$, the main determinant of disc radius `now' (i.e. time
$t_{uv}$ after switch on) is the initial disc mass. This can be
simply understood in that a disc that is initially more extended
will suffer stronger photoevaporative mass loss and therefore
will also shrink much faster: the convergence of the models shown
in Figure 4 demonstrates the same effect. 

  We therefore conclude that the fact that  $< 20 \%$ of ionised
flows contain discs with radii $> 50$ A.U. is a fact that chiefly
constrains the initial {\it masses} of circumstellar discs, rather
than their initial radii. From our models, the predominance of
`small' (less than $50$ A.U.) discs simply translates into
a predominance of discs with initial masses $< 0.1 M_\odot$ (if
$t_{uv} = 1.5$ Myr) or with initial masses $ < 0.03 M_\odot$ (if
$t_{uv} = 1$ Myr). This predominance of smaller mass
discs is at least compatible with the submillimetre results of
Andrews and Williams (2005) for the {\it current} (as opposed
to initial) masses of discs in Taurus Auriga.  

\section {Discussion}

Our models have shown that the common occurrence of stars with
optically thick {\it inner} discs in the ONC implies that the
ultraviolet source (presumably $\theta_1 $C) switched on less
than $1-2$ Myr ago, and that there has been little star formation
since that time. In the alternative scenario (where stars form
continuously in the presence of a constant ultraviolet field), the 
preponderance of small discs (radius $< 50$ A.U.) in the observed
distribution would imply that discs were formed with
implausibly low masses (much less than the minimum mass solar
nebula).

 Given the scenario that discs are exposed to the ultraviolet
field for a finite time, the main parameter that affects their
present day size distribution is the distribution of initial
disc {\it masses}. The present day size distribution is in fact
remarkably insensitive to the initial distribution of disc radius:
at a given disc mass, an extended disc suffers stronger photoevaporative
mass loss and so shrinks to radii similar to discs of much smaller
radius initially.

The relative scarcity of discs larger than $50$ A.U. imposes upper limits on the initial disc masses for the bulk of the population in the range
$0.03-0.1 M_\odot$. 
These upper mass limits are in no way
unexpected; what is possibly more interesting is the fact that
$20 \%$ of ionised flows do contain larger discs, and therefore
must have larger initial disc masses. In fact, the initial
disc masses that are required in order to produce the largest 
(i.e. resolved) discs are in the range where disc self-gravity was almost
certainly important. [Recall that a measure of the importance
of disc self-gravity is provided by the Toomre $Q$ criterion, and can
be re-expressed as the approximate condition $M_d/M_* > H/R$. Since
$H/R$ (the aspect ratio of the disc) is typically $\sim 10 \%$ in
models of circumstellar discs, the disc masses that we are invoking
as `initial' conditions for the resolved disc population are clearly
close to, or in, the self-gravitating, regime.]
The fact that $\sim 20 \%$ of the discs were probably self-gravitating
at the time that the ultraviolet source switched on implies that at least
$20 \%$ of discs pass through a self-gravitating phase and would also
be consistent with all discs spending $20 \%$ of their lifetimes in the
self-gravitating state. However, according to our models, the 
resolved discs are {\it not} self-gravitating now (i.e. after exposure to
the ultraviolet field for $1-2$ Myr) since they have been significantly
eroded in the meantime. 

  How do these results affect an understanding of the so-called
proplyd lifetime problem? This problem stems from estimating
the exhaustion timescale ($t_{exh}$) for a disc as simply the ratio of the
disc mass to the photoevaporation rate. Since estimates of disc masses
have (until the recent work of Eisner and Carpenter 2006: see below) 
been uniformly low  
(Mundy et al  1995, Bally et al 1998b) whereas mass loss
rates (both from modeling and from spectroscopic measurements;
Henney \& O'Dell 1999) are high, the resulting values of 
$t_{exh}$ are low ($< 10^5$ years), i.e. much less than the mean
stellar age in the ONC ($> 1$ Myr; Palla and Stahler 1999). This
has been interpreted as implying a requirement that discs have
been only very recently exposed to the ultraviolet flux of the central
OB stars: either because their orbits only take them briefly into
the cluster core (St\"orzer and Hollenbach 1999) or due to the recent
switch on of the ionising stars. However, Scally and Clarke (2001)
argued against the former hypothesis on the grounds that it proved
impossible to find plausible dynamical models of the ONC whose
orbital structure could solve the proplyd lifetime problem.

 The requirement that the discs in the ONC are only exposed to the
ultraviolet radiation field of the central stars over a timescale
$< 10^5$ years is much more stringent than what we have found in our
models, which can readily accommodate the exposure of discs for
a Myr or more. 
 In order to understand this, we must look in more
detail at the model predictions: the bulk of the population
in the models is composed of discs that are compact and low
in mass, with correspondingly low photoevaporative mass loss
rates (see Figure 1). On the other hand, the large (resolved)
discs in the models have both higher masses and higher photoevaporative
mass loss rates. In both types of model systems, $t_{exh}$ is
relatively long ($\sim$ a Myr). 

 The reason that the models do not
exhibit a `proplyd lifetime problem' is thus chiefly because 
those models 
that spend a reasonable time with large disc radii ($\sim 100$ A.U.)
have correspondingly
large disc masses at that stage ($\sim 0.1 M_\odot$). At first sight,
this would appear to contradict the low disc masses (and upper limits)
obtained from submillimetre studies of proplyd discs. Disc
masses of around $0.02 M_\odot$ have been measured in a handful of
proplyd systems (i.e. 182-413 (HST 10) by  Bally et al 1998b) and 
163-317, 170-337,171-334 and 171-340 by Williams et al 2005).
Crucially, these studies assumed {\it optically thin} emission.
Thus, using the opacities assumed by these authors, one may calculate
a lower limit to the emitting area of the  disc, in order for the
assumption of optically thin emission to be correct. 
It is however notable that in the subset of systems for which the
disc size and inclination is known through optical imaging
(i.e. 182-413, 170-337 and 171-340), the disc emitting area
is remarkably close to this lower limit: in other words, the observed
fluxes are close to that expected for optically thick emission
for discs of this size scale. In this case, the observed fluxes
only impose a {\it lower limit} on the disc mass. This surmise
is supported by the recent detection at 3mm of a relatively massive
($\sim 0.13 M_\odot$) disc in 182-413, as well as in a handful
of other proplyds (Eisner and Carpenter 2006). The mean disc
masses inferred in undetected sources is likewise higher in the
case of the 3mm measurements than those derived from submillimetre
observations, which is again compatible with the notion that some
of the sources may be optically thick in the latter case.
A possible way of exploring this issue further would be to concentrate
millimetre studies on systems known to harbour  more extended discs (e.g. those
detailed by Vicente and Alves 2005).

\section{Conclusions}

Our modeling of the photoevaporation of viscously evolving discs
in the Orion Nebula Cluster, combined with observational
data on the observed size distribution of discs in the cluster,
has allowed us to draw a number of
conclusions about the properties of the discs and the history of
irradiation by the cluster's dominant OB star ($\theta_1 C$).

  1) The fact that most stars in the core of the ONC possess discs
but that the majority ($> 80 \%$) of such discs 
are compact ($< 50$ A.U. in radius) is {\it incompatible} with
a model in which star-disc systems are continuously created in the
steady ultraviolet field of $\theta_1 C$. Instead, we require that
$\theta_1 C$ has been `switched on' for no more than $1-2$ Myr, in which
case the bulk of discs have been shorn by prior photoevaporation to
their present day compact state.
 
 2) We find that the present day sizes of discs that are subject
to a mixture of photoevaporation and viscous evolution are  very insensitive
to their initial sizes. This is because   discs that are
more extended initially suffer stronger photoevaporative mass loss
and thus shrink back to sizes similar to systems that were initially
much more compact. The main parameter  affecting present day disc sizes 
is the  disc mass at the stage that the photoionising source switched on.

 3) The bulk of discs in the ONC, which are not resolved by HST,
have radii less than $20-50$ A.U. and must, in our models, correspond
to systems with disc masses, when the photoionising source
switched on, of less than $0.03-0.1 M_\odot$ ( the
range in these upper limits depending on the duration of exposure
to the photoionising field).

4) The minority (around $20 \%$) of  discs in the ONC that are
more extended (i.e. with radius $> 20-50$ A.U. and hence resolvable
by HST) correspond to a population of more massive discs 
at the stage that the photoionising source switched on. Such
discs were almost certianly self-gravitating at this stage, but,
following $1-2$ Myr of photoevaporation, would not be expected to be
self-gravitating currently.   

5) In our models,   discs are photoevaporated on a timecsale
of $\sim 10^6$ years and so do not imply any conflict with the
existence of discs in the ONC. We suggest that the much shorter,
observationally based, 
photoevaporation timescales quoted in the literature (which are
associated with the `proplyd lifetime problem') may result from
an under-estimate of disc masses in systems that are optically
thick at submillimetre wavelengths. In order to avoid ambiguities
due to optical depth, it is desirable that future measurements
concentrate on proplyds containing the most 
spatially extended discs.

\section{Acknowledgments}
I am grateful to Jens Rodmann and Mark McCaughrean for sharing
data with me in advance of publication and for useful discussions.
I also acknowledge useful input from Richard Alexander,   Josh
Eisner and the referee, Ian Bonnell.
 



\begin{thebibliography}{}

\bibitem{} Adams, F.C., Hollenbach, D., Laughlin, G., Gorti. U., 2004, ApJ 611,360

\bibitem {} Alexander, R., Armitage, P., 2006. ApJ  639, L83

\bibitem {} Alexander, R., Armitage, P., 2007. MNRAS in press. 

\bibitem{} Andrews, S.M., Williams, J.P., 2005. ApJ 631,1134

\bibitem{} Andrews, S.M., Williams, J.P., 2007. ApJ submitted (astro-ph /0610813). 
\bibitem{} Armitage,  P.J., Clarke, C.J., 1997, MNRAS 285,540

\bibitem{} Armitage,  P.J., Clarke, C.J., Palla, F., 2003,
MNRAS 342,1139

\bibitem{} Bally, J., O'Dell, C.R., McCaughrean, M., 2000, AJ 119,2919

\bibitem{} Bally, J., Sutherland, R., Devine et al 1998a), AJ 116,293

\bibitem{} Bally, J., Testi, L., Sargent, A., Carlstrom, J., 1998b), AJ 116,854

\bibitem{} Bate, M.R., Bonnell, I.A., Bromm, V., 2003. MNRAS 339,577

\bibitem{} Burkert, A., Bodenheimer, P., 2000, ApJ 543,822

\bibitem{} Clarke, C.J., Gendrin, A. and Sotomayor, M., 2001, MNRAS 328,485

\bibitem{} Dale, J., Bonnell, I., Clarke, C., Bate, M., 2005. MNRAS 358,291

\bibitem{} Dullemond, C., Natta, A., Testi, L., 2006. ApJ 645,L69
 
\bibitem{} Eisner, J., Carpenter, J., 2006. ApJ 641,1162

\bibitem{} Hartmann, L., Calvet. N., Gullbring, E., D'Alessio, P., 1998,
ApJ 495,385

\bibitem{} Henney, W.J., Arthur, S.J., 1998. AJ 116,322

\bibitem{} Henney, W.J., O'Dell, C.R., 1999. AJ 118,2350

\bibitem{} Johnstone, D., Hollenbach, D., Bally, J., 1998. ApJ 499,758

\bibitem{} Kitamura, Y., Momose, M., Yokogawa, S., Kawabe, R., Tamura, M.,
Ida, S., 2002. ApJ 581,357
  
\bibitem{} Lada, C.J., Muench, A. A., Haisch, K.E., Lada, E.A., Alves, J.F., Tollestrup, E.V., Willner, S.P., 2000, AJ 120,3162


\bibitem{} Lynden-Bell, D., Pringle, J.E., 1974. MNRAS 168,603

\bibitem{} McCaughrean, M.J., O'Dell, C.R., 1996. AJ 111,1977
\bibitem{} Miyake, K., Nakagawa, Y., 1993. Icaru, 106,20

\bibitem{} Mundy, L.G., Looney, L.W., Lada, E.A., 1995, ApJ 452,137

\bibitem{} O'Dell, C.R., Wen, Z., Hu, X., 1993. ApJ 410,696

\bibitem{} O'Dell, C.R., Wong, K., 1996. AJ 111,846

\bibitem{} Palla, F., Stahler, S.W., 1999, ApJ 525,772

\bibitem{} Rodmann, J., 2002.  Diploma Thesis, Univ. Potsdam, Germany


\bibitem{} Scally, A., Clarke, C.J., 2001, MNRAS 325,449

\bibitem{} Shakura, N.I., Sunyaev, R.A., 1973. A\&A 24,337


\bibitem{} St\"orzer, H., Hollenbach, D., 1999. ApJ 515,669

\bibitem{} Takeuchi, T., Clarke, C., Lin, D., 2005. ApJ 627, 286

\bibitem{} Throop, H., Bally, J., 2005. ApJ 623,L149

\bibitem{} Vicente, S.M., Alves, J., 2005. astro-ph 0506585

\bibitem{} Williams, J., Andrews, S., Wilner, D., 2005. ApJ 634,495

\bibitem{} Wilson, T.L., Filges, L., Codella, C., Reich, W., Reich, P., 1997. A \& A 327,1177

\end{thebibliography}
\end{document}